\documentstyle[aps,epsf,twocolumn]{revtex}
\def\la{\langle} 
\def\ra{\rangle} 
\def\be{\begin{eqnarray}} 
\def\ee{\end{eqnarray}}

\newcommand{\eq}{\begin{equation}} \newcommand{\eqx}{\end{equation}}

\newcommand{\eqn}{\begin{eqnarray}} \newcommand{\eqnx}{\end{eqnarray}}

\begin{document}
\draft

\title{\bf Chiral Disorder and QCD at Finite Chemical Potential}

\author{ {\bf Romuald A. Janik}$^1$, {\bf Maciej A.  Nowak}$^{1}$ ,
{\bf G\'{a}bor Papp}$^{2}$ and {\bf Ismail Zahed}$^3$}

\address{$^1$ Department of Physics, Jagellonian University, 30-059
Krakow, Poland.
\\ $^2$ITP, Univ. Heidelberg, Philosophenweg 19, D-69120 Heidelberg, 
        Germany \& \\ Institute for
Theoretical Physics, E\"{o}tv\"{o}s University, Budapest, Hungary\\
$^3$Department of Physics and Astronomy, SUNY, Stony Brook, 
New York 11794, USA.}
\date{\today} \maketitle

\begin{abstract}
We investigate the effects of a finite chemical potential $\mu$
in QCD viewed as a disordered medium. In the quenched approximation,
$A_4=i\mu$ induces a complex electric Aharonov-Bohm effect that
causes the diagonal contribution to the quark return probability to 
vanish at $\mu=m_{\pi}/2$ (half the pion mass). In two-color QCD,
the weak-localization contribution to the quark return probability
remains unaffected causing a mutation in the spectral statistics.
In full QCD, the complex
electric flux is screened and the light quarks are shown to diffuse 
asymmetrically with a substantial decrease in the conductivity along 
the `spatial' directions. Mean-field arguments suggest that a d=1 
percolation transition may take place in the range $1.5\rho_0<\rho<3\rho_0$, 
where $\rho_0$ is nuclear matter density.

\end{abstract}

{\bf 1.}
QCD at finite chemical potential $\mu$ is still not well understood
despite the many efforts invested by a number of groups in the past
years.  Lattice Monte-Carlo algorithms are difficult to 
implement at finite $\mu$ owing to the complex character of the 
measure~\cite{QCD}. Results from strong coupling arguments~\cite{STRONG} 
and quenched simulations~\cite{QCD,QUENCHED} are available but 
do not seem to be transparent physically. A number of theoretical 
constraints can be implemented using symmetry and data~\cite{BOOK}.
However, they are only reliable for densities typically of the order of
nuclear matter density $\rho_0$. Results from constituent quark models 
at higher densities point at the possibility of a chiral transition 
at about 3 times nuclear matter density~\cite{NJL} and the occurrence
of a diquark superconducting phase at even higher densities~\cite{SUPER}. 

In this letter we would like to address the effects of a finite chemical 
potential on the chiral disorder of the QCD ground state. At $\mu=0$ we 
have recently~\cite{USPRL,US} shown that light quarks in a finite
Euclidean volume $V$ 
are in a diffusive mode, with a diffusion constant $D=2F^2/\Sigma$
where $F$ is the weak pion decay constant and $\Sigma=|\la\overline{q} q\ra|$
the light quark condensate. The effects of matter cause the medium to 
change thereby affecting the diffusion properties of the light quarks.
In many ways our problem is similar to the problem of electrons in disordered
metals under the influence of external sources~\cite{MONTAMBAUX}.

\vskip 0.3cm

{\bf 2.}
The eigenvalue equation of the Dirac operator for fundamental quarks in
a fixed gluon field $A$ at finite chemical potential $\mu$ is
\be
\left( i\nabla \!\!\!\!/[A] +i\mu\gamma_4 \right) \,q_k =\lambda_k [A] \, q_k \,\,.
\label{01}
\ee
for the right-eigenfunctions, and 
\be
\left( i\nabla \!\!\!\!/ [A] -i\mu\gamma_4 \right) \,Q_k =\lambda_k^* [A] \, Q_k \,\,.
\label{001}
\ee
for the left-eigenfunctions. The eigenvalues are complex and paired by chiral 
symmetry. The  set $(q_k, Q_k)$ is biorthogonal.
Generalizing the construction~\cite{USPRL} for  the case of a finite chemical 
potential, we may write the probability $p(t,\mu )$ for a light quark
 to start at $x(0)$
in $V$ and return back to the same position $x(t)$ after a proper time
duration $t$, as
\be
p(t, \mu )= \frac {V^2}N\,e^{-2m |t|}
\Big\la |\la x(0)|e^{i(i\nabla \!\!\!\!/ [A] +i\mu\gamma_4)|t|}|x(0)\ra|^2\Big\ra_A\,.
\label{1}
\ee
The averaging in (\ref{1}) is over all gluon 
configurations using the unquenched QCD measure with massive (sea) quarks.
The normalization in (\ref{1}) is per state, where $N$ is the total number
of quark states in the four-volume $V$. Equation~(\ref{1}) can be resolved in terms of
(\ref{01}-\ref{001})
\be
&&p(t) =\frac {V^2}{N} e^{-2m|t|} \sum_{j,k} \nonumber\\
&&\times\Big\la e^{i|t| (\lambda_j-\lambda_k^*)[A]} 
q_j (x)Q_j^* (x) \, Q_k (x)q_k^* (x) \Big\ra_A
\label{02}
\ee
where the exponent $e^{-2m|t|}$ is solely due to the valence quark mass.
We note that (\ref{02}) is gauge-invariant and amenable to 
lattice Monte-Carlo simulation. It requires both the eigenvalues
and eigenfunctions.

For analytical considerations, it is best to rewrite (\ref{1}) in terms
of the standard Euclidean propagators for the quark field,
\be
p(t, \mu ) = \frac {V^2}N \lim_{y\to x}{}&&
\int \frac {d\lambda_1d\lambda_2}{(2\pi)^2} 
\,e^{-i(\lambda_1-\lambda_2) |t|}\nonumber\\&&
\times\Big\la {\rm Tr}\left( S(x,y;z_1,\mu) S^{\dagger} (x,y; z_2,\mu)\right)\Big\ra_A
\label{des2}
\ee
with $z_{1,2}=m-i\lambda_{1,2}$, and
\be
S(x,y; z, \mu) = \la x| \frac 1{i\nabla  \!\!\!\!/[A] + i\mu\gamma_4 + iz} |y\ra \,.
\label{des3}
\ee
Since the eigenvalues (\ref{01}) are complex, it is important that
$m>{\rm max\,\, Im}\lambda_k$ in (\ref{des2}-\ref{des3}). 
For small $\mu$ the imaginary
parts are of order $\mu^2$ (second order perturbation theory) so it is
enough to have $m>\mu^2$ in units where the infrared scale is of order 1. 
For large $\mu$, $m$ should be made large and then reduced after integration.
This will be understood throughout.

Setting $\lambda_{1,2}=\Lambda\pm \lambda/2$ and neglecting the effects
of $\Lambda$ in the averaging in (\ref{des2})~\cite{USPRL},
we find that in the flavor symmetric
limit, the correlation function in (\ref{des2}) relates to the 
`baryonic' pion correlation function~\cite{LOMBARDO}
after a proper analytical continuation 
of the  current quark mass~\cite{USPRL}. Specifically,
\be
p(t, \mu) = \frac {EV^2}{2\pi N} \lim_{y\to x}{}
\int \frac {d\lambda}{2\pi} 
\,e^{-i\lambda |t|} {\bf C}_{\pi_B} (x,y; z)
\label{des55}
\ee
where
\be
&&\,{\bf 1}^{ab}\, {\bf C}_{\pi_B} (x,y;z) = \nonumber\\
&&\Big\la {\rm Tr}\left(
S(x, y;z,\mu) i\gamma_5\tau^a S(y,x;z,-\mu) i\gamma_5\tau^b\right)\Big\ra_A
\label{des5}
\ee
with $z=m-i\lambda/2$ and $E=\int d\Lambda$. For conventional pions both
propagators in (\ref{des5}) carry $\mu$ with the same sign.

\vskip 0.3cm
{\bf 3.}
The effects of $\mu$ in (\ref{01}) is that of a complex and constant
4-vector potential
$A_4=i\mu$. It breaks particle-antiparticle symmetry much like a vector
potential breaks particle-particle (antiparticle-antiparticle) symmetry.
It acts like a complex electric Aharonov-Bohm  effect in the 
particle-antiparticle channel. The particle-antiparticle system breaks apart
for $\mu$ typically of the order of the binding energy (about the pion mass). 
This phenomenon is reminiscent of the destruction of heavy-mesons by 
chromo-electric fields~\cite{ADAMI}, and charge or spin density waves by 
transverse electric fields~\cite{SPIN}, although not identical since 
in our case the `electric field' is zero.

In the quenched approximation, the only dependence on $\mu$ in (\ref{des5})
is that shown in the external propagators. In the semi-classical approximation
and for three colors (for two colors see below) it acts as a `complex'
flux on the `diffusons' (particle-antiparticle)~\cite{USPRL}. 
For $z=m$, the long paths contributions to (\ref{des5}) are given by
\be
{\bf C}_{\pi_B} (x,y;m) \approx \frac 1V \sum_Q e^{iQ\cdot (x-y)} 
\frac {\Sigma^2}{F^2} \frac 1{\tilde{Q}^2+m_{\pi}^2}
\label{des06}
\ee
with $Q_{\alpha} =n_{\alpha}2\pi/L$ and
 $\tilde{Q}_{\alpha} = Q_{\alpha} +2i\mu\delta_{4\alpha}$ 
in $V=L^4$. The factor 2 in front of $A_4$ reflects on the fact that
the fluxes {\it add} in the `diffuson'. 
Using the Gell-Mann Oakes Renner (GOR) relation $F^2m_{\pi}^2=m\Sigma$, 
and the analytical continuation $m\rightarrow m-i\lambda/2$, we 
find
\be
{\bf C}_{\pi_B} (x,y;z) \approx \frac 1V \sum_Q e^{iQ\cdot (x-y)} 
\frac {2\Sigma}{-i\lambda + 2m + D\tilde{Q}^2}
\label{des6}
\ee
with the diffusion constant $D=2F^2/\Sigma$~\cite{USPRL}. 
Inserting (\ref{des6}) into (\ref{des55}), we
observe that the `diffusion' pole in the lower
part of the complex plane depends critically
on the value of the chemical potential $\mu$.

In the zero mode approximation $n_{\alpha}=0$ or for large times $t>\tau_{\rm 
erg}=L^2/D$, the quark return probability is
\be
p(t, \mu) \approx \theta(m_{\pi}-2\mu) \,\,e^{-D(m_{\pi}^2-4\mu^2)|t|}
\label{des7}
\ee
where we have used $E/\Delta=N$ and $\varrho=1/\Delta V$, with 
$\Sigma=\pi \varrho$, according to the Banks-Casher relation.
 Here $\Delta$ is the mean interlevel spacing
between 
the eigenvalues for $\mu=0$. The occurrence of the step-function theta
in (\ref{des7}) reflects on the fact that the diffusion pole moves
from the lower-half to the upper-half of the complex $\lambda$-plane.
For $t>\tau_{\rm erg}$ the quark return probability vanishes for
$\mu=m_{\pi}/2$ in the quenched approximation. Physically, this
means that the complex electric flux splits the quark-antiquark pair
in the quenched approximation a situation reminiscent of the
magnetic fluxes in type-I superconductors~\cite{SUPER1}. This result is 
consistent with current quenched lattice simulations~\cite{QUENCHED}
and the results of schematic chiral random matrix models for finite
$\mu$~\cite{STEPHANOV,USMUX,THEMMUX}.

We note that in the double scaling limit $D m_{\pi}^2 t_H\sim mV\ll 1$
and $D\mu^2 t_H\sim \mu^2 V\ll 1$, (\ref{des7}) is about 1 and universal.
This regime is amenable to a random matrix model analysis and signals
the onset of a new universality for the complex eigenvalues of (\ref{01}).
It can be modeled using a reduction to 0-dimension.

\vskip 0.3cm

{\bf 4.}
For two-color QCD the situation is special since in this case
the Dirac operator possesses an additional symmetry~\cite{LEUTSMILGA} due to 
the pseudo-real nature  of the SU(2) representations.  In the diffusive picture
of the QCD vacuum this means that the quark return probability
receives contributions from both `diffusons' (diagonal) and `cooperons' 
(interference) paths in the semi-classical approximation~\cite{USPRL}
(and references therein). The `cooperons' are just the weak-localization
contribution to the quark return probability~\cite{MONTAMBAUX}. In the standard
description of diffusion they follow from the interference between the
classical loops traveled in opposite directions (coherent backscattering)
and reflect on the time-reversal invariance of the underlying microscopic 
Hamiltonian. While the `diffusons' sense 2 flux lines, the `cooperons' are 
flux-blind. A rerun of the above arguments now give 
\be
{\bf C}_{\pi_B} (x,y;z) \approx&&+ \frac 1V \sum_Q e^{iQ \cdot (x-y)} 
\frac {2\Sigma}{D\tilde{Q}^2+2m-i\lambda}\nonumber\\
&&+\frac 1V \sum_Q e^{i{Q}\cdot (x-y)} 
\frac {2\Sigma}{D{Q}^2+2m-i\lambda}
\label{two1}
\ee
instead of (\ref{des06}). The
first term is the `diffuson' contribution, 
and the second term the `cooperon' contribution. Inserting (\ref{two1}) into
(\ref{des55}) yields
\be
p(t, \mu) \approx e^{-Dm_{\pi}^2 |t|}\left(
\theta(m_{\pi}-2\mu) \,\,e^{+4D\mu^2|t|} +1\right)
\label{dess7}
\ee
in the zero mode approximation or $t>\tau_{\rm erg}=L^2/D$. 
For $\mu>m_{\pi}/2$ the `diffuson' contribution (first term)
drops and we are only left with the `cooperon'
contribution which is of order $e^{-2m |t|}$. The 
latter is of order 1 and universal for $\mu^2<m\ll 1/V$.
The transition to the `cooperon' phase is simply a transition
to the superconducting phase in this case.

In the universal regime and for small current quark masses, the chemical
potential is $\mu\ll 1/\sqrt{V}$ and small. Hence, the complex eigenvalues
$\lambda_k$ carry an imaginary part of order $1/V$ which is of the order of
the microscopic level spacing for the `unperturbed' real parts. If we focus
on the level-correlations between only the real parts of $\lambda$'s in 
the microscopic limit $x=V\lambda\sim 1$ we expect a mutation in the
level correlations from the orthogonal to unitary ensemble. The mutation
follows a migration of part of the quark levels from the real axis to the 
complex plane under the influence of the tiny chemical potential. The spectral 
rigidity  $\Sigma_2 (N,\mu)$ for the real parts of $\lambda$'s can be 
estimated using semi-classical arguments~\cite{USPRL}. The result is 
\cite{NOTE5}
\be
\Sigma_2 (N, \mu)\approx &&\theta (m_{\pi}-2\mu) \,
\frac 1{2\pi^2}{\rm ln} \left( 1+\frac 
{N^2}{\tilde{\alpha}^2}\right)\nonumber\\
&&+\frac 1{2\pi^2}{\rm ln} \left( 1+\frac 
{N^2}{{\alpha}^2}\right)
\label{rigid}
\ee
where $N=E/\Delta\gg 1$, 
$\tilde{\alpha}=D(m_{\pi}^2-4\mu^2)/2\Delta$ and $\alpha=2m/\Delta$.
The level spacing $\Delta=1/\varrho V$ is taken to be that of the $\mu=0$ limit. 
The behavior (\ref{rigid}) can be addressed using current quenched lattice 
Monte-Carlo simulations in QCD~\cite{QCD,QUENCHED}.

\vskip 0.3cm

{\bf 5.}
In unquenched QCD, the vacuum supports quark-antiquark pairs.
The `baryonic' pion correlations are screened by pair creation,
rendering the quark-antiquark system blind to the complex electric
Aharonov-Bohm flux (constant $A_4$). As a result, the correlations
in (\ref{des5}) are primarily that of a quark-antiquark in a vacuum 
for zero nucleon density with $\mu\leq m_N/3$ where $m_N$ is the 
nucleon mass (ignoring binding energies).

At finite nucleon density,
the quark return probability follows from a pertinent analytical continuation 
of the pion propagator in matter. In a mean-field approximation we 
have~\cite{WIRZBA,NOTE6}
\be
&&{\bf C}_{\pi_B} (x,y;m) \approx \frac 1V \sum_Q e^{i{Q}\cdot (x-y)} 
\frac {(1-\alpha\rho)^2}{(1-\beta\rho)}
\frac {\Sigma^2}{F^2} \nonumber\\
&&\times\left( Q_4^2+ \left(\frac {1-\gamma\rho}{1-\beta\rho}\right)\vec{Q}^2
+ \left(\frac {1-\alpha\rho}{1-\beta\rho}\right)  m_{\pi}^2\right)^{-1}
\label{f1}
\ee
with ${Q}_{\sigma} =n_{\sigma}2\pi/L$ in $V=L^4$.
Here $\alpha=\la N|\overline{q}q|N\ra/\Sigma$ measures the  strength of
the pion-nucleon sigma term relative to the scalar condensate, with 
$1/\alpha\sim 3\rho_0$. The parameters  $\beta\sim \alpha$ and 
$\gamma\sim 2\alpha$ relate to the S-wave pion-nucleon scattering 
lengths~\cite{BOOK,WIRZBA}. The leading density approximation follows from the 
mean-field analysis by keeping only the leading term in the nucleon density 
$\rho$~\cite{BOOK,WIRZBA}. In the space-like regime under consideration 
there is no imaginary contribution to (\ref{f1}).

Using the GOR relation $F^2m_{\pi}^2=m\Sigma$, 
and the analytical continuation $m\rightarrow m-i\lambda/2$, we 
may rewrite (\ref{f1}) as
\be
{\bf C}_{\pi_B} (x,y;z) \approx &&\frac 1V \sum_Q e^{i{Q}\cdot (x-y)} 
\nonumber\\&&\times
\frac {2\Sigma \, (1-\alpha \rho)}{-i\lambda + 2m + D_4{Q}_4^2
+D_S \vec{Q}^2}
\label{f2}
\ee
with temporal and spatial diffusion coefficients
\be
&&D_4= D\,\frac {1-\beta\rho}{1-\alpha\rho}, \nonumber\\
&&D_S= D\,\frac {1-\gamma\rho}{1-\alpha\rho} \,.
\label{f3}
\ee
Hence $D_4\sim D$ and $\rho$ independent, while $D_S$ vanishes
for $\rho\sim 1/\gamma\sim 1/(2\alpha )\sim 1.5 \rho_0$ in the mean-field
approximation~\cite{NOTE}. At this point the quark density of states
at zero virtuality is about $\varrho (\mu)\sim (1-\alpha \rho)\Sigma\sim
\varrho/2 $ by the Banks-Casher relation~\cite{BANKS}. Using the
Kubo-formula we conclude that the conductivity vanishes along the spatial
directions $\sigma_S=D_S\varrho =0$. This is not a metal-insulator
transition as the conductivity $\sigma_4=D_4\varrho\neq 0$ is still 
non-zero.  It can be regarded as an `asymmetric' percolation transition
from d=4 to d=1, with a diffusive quark return probability~\cite{US} of
the form 
\be
p(t, \mu)\approx \frac{e^{-2m|t|}}{\sqrt{4\pi E_4 |t|}}
\label{returnx}
\ee
where we have used $E/\Delta_* =N\gg 1$ and a density dependent
level spacing $\Delta_* /\Delta\sim 1/(1-\alpha\rho)$. 
Here $E_4=D_4/L^2$ is the `temporal' Thouless energy as opposed to
$E_S=D_S/L^2$ the `spatial' Thouless energy. A similar phenomenon 
takes place at finite temperature~\cite{US}.

At this stage there are two courses of action: The conductivity
$\sigma_4$ vanishes continuously with the depletion of the number
of quark states at zero virtuality corresponding to a vanishing
of the quark density of states at $\rho\sim 3\rho_0$.
(In fact this is what happens if only the leading density approximation
were used.) This transition is likely of
second-order or higher and would be in overall
agreement with some constituent quark model results~\cite{NJL}. 
Alternatively,
it may terminate abruptly for a density  $1.5\rho_0<\rho<3\rho_0$ through
a d=1 percolation transition. This is intuitively more appealing if
we were to proceed from the high-density region backward, and support the
idea that a nucleon Fermi-surface in 3-dimensions correspond to an array 
of `rods' in d=4 Euclidean space (nucleon-worldlines) forcing the conductivity
to be essentially
1-dimensional by Pauli-blocking. Ideas in favor of a percolation transition 
at finite density have been also stressed recently by Satz~\cite{SATZ} using
different arguments.

\vskip 0.3cm

{\bf 6.} 
We have shown that the disordered properties of the QCD ground state are
quantitatively altered by a finite chemical potential $\mu$. In quenched
QCD the effects of $A_4=i\mu$ are analogous to that of a complex
electric Aharonov-Bohm effect, causing the `baryonic'
quark-antiquark pair to accumulate 2 flux lines and 
rupture at $\mu=m_{\pi}/2$. This result is in agreement with quenched lattice 
simulations~\cite{QCD,QUENCHED}. In two-color QCD, the quark-quark and
antiquark-antiquark pairs
are flux-blind. As a result, the weak-localization contribution to the
quark return probability remains unaffected. In the universal limit $\mu^2<m\ll 
1/V$, the quark spectrum in two-color QCD exhibits a change in
the spectral statistics from an orthogonal to unitary ensemble
again at $\mu=m_{\pi}/2$. 

In the unquenched approximation the electric flux is screened and the
diffusion becomes asymmetric. The light quarks take longer time to diffuse 
along the spatial directions owing to the presence of a Fermi surface. This
asymmetry is commensurate with the softening in the pion dispersion 
relation at $\rho\sim \rho_0$ due to pion-nucleon S-wave rescattering. 
As a result, the bulk Ohmic conductivity of the system becomes quasi 
1-dimensional, with a potential for a d=1 percolation transition in the density
range $1.5\rho_0<\rho<3\rho_0$.

Most of our results can be numerically 
checked by analyzing the  quark return probability in the ergodic and
diffusive regime at finite chemical potential in lattice QCD or in continuum 
models such as the instanton liquid model~\cite{BOOK,INST}.

\vskip 0.5cm
{\bf Acknowledgements}
\\
IZ would like to thank Edward Shuryak for a discussion.
This work was supported in part by the US DOE grant DE-FG-88ER40388, by the 
Polish Government Project (KBN) grants 2P03B04412 and 2P03B00814 and by the 
Hungarian grants FKFP-0126/1997 and OTKA-F026622.


\begin{thebibliography}{99}
\bibitem{NOTE5}
Our normalization of the spectral form factor in \cite{USPRL}
is such that the spectral rigidity for the orthogonal
ensemble is recovered at $\mu =0$.
\bibitem{NOTE6}
In the vacuum, the baryonic pion couples
in the same way to the nucleon as the standard pion. 
In matter, there may be extra couplings
which we are ignoring for simplicity. They can be
addressed using general principles and symmetry.
\end{thebibliography}
\end{document}